\documentclass[aps,prl,twocolumn,floats,showpacs,floatfix,a4paper]{revtex4}

\usepackage{epsfig}
\usepackage{color}
\usepackage{bm}
\usepackage{latexsym}
\usepackage[utf8x]{inputenc}
\usepackage{ucs}

\begin{document}

\newcommand{\mm}[1]{{\mathbf{#1}}}
\newcommand{\cc}{{\bf\Large C }}
\newcommand{\hide}[1]{}
\newcommand{\tbox}[1]{\mbox{\tiny #1}}
\newcommand{\half}{\mbox{\small $\frac{1}{2}$}}
\newcommand{\sinc}{\mbox{sinc}}
\newcommand{\const}{\mbox{const}}
\newcommand{\tr}{\mbox{tr}}
\newcommand{\intt}{\int\!\!\!\!\int }
\newcommand{\ointt}{\int\!\!\!\!\int\!\!\!\!\!\circ\ }
\newcommand{\eexp}{\mbox{e}^}
\newcommand{\EPS} {\mbox{\LARGE $\epsilon$}}
\newcommand{\ar}{\mathsf r}
\newcommand{\im}{{\cal I}m}
\newcommand{\re}{{\cal R}e}
\newcommand{\bmsf}[1]{\bm{\mathsf{#1}}}
\newcommand{\dd}[1]{\:\mbox{d}#1}
\newcommand{\abs}[1]{\left|#1\right|}
\newcommand{\bra}[1]{\left\langle #1\right|}
\newcommand{\ket}[1]{\left|#1\right\rangle }
\newcommand{\mbf}[1]{{\mathbf #1}}
\newcommand{\eos}{\,.}
\definecolor{red}{rgb}{1,0.0,0.0}

\title{${\cal PT}$-symmetry in macroscopic magnetic structures}

\author{J. M. Lee, T. Kottos}
\affiliation{Department of Physics, Wesleyan University, Middletown, Connecticut 06459}
\author{B. Shapiro}
\affiliation{Technion - Israel Institute of Technology, Technion City, Haifa 32000, Israel}
\date{\today}

\begin{abstract}
We introduce the notion of ${\cal PT}$-symmetry in magnetic nanostructures and show that they can 
support a new type of non-Hermitian dynamics. Using the simplest possible set-up consisting of two 
coupled ferromagnetic films, one with loss and another one with a balanced amount of gain, we demonstrate the 
existence of a spontaneous ${\cal PT}$-symmetry breaking point where both the eigenfrequencies and eigenvectors are 
degenerate. Below this point the frequency spectrum is real indicating stable dynamics while above this point it is complex 
signaling unstable dynamics. 
\end{abstract}

\pacs{11.30.Er, 76.50.+g, 05.45.Xt,}
\maketitle

Spin dynamics in synthetic magnetic nanostructures has attracted increasing attention during the last years \cite{GM96}, 
because of  the interesting fundamental physics involved and also due to its important practical implications: 
magnetic storage and information processing \cite{A05,SHKKTIR07}, sensing \cite{GDJ08}, and creation of tunable high 
frequency oscillators \cite{RS08} are some of the areas that have been benefited by this research activity. An important 
step in this endeavor is the realization of new magnetic nanodevice architectures with additional degrees of freedom 
which permit better control of magnetization dynamics. 

Along the same lines, management of classical wave propagation via synthetic structures has been proven to be 
successful resulting in the creation of new materials with unexpected properties. Examples of this success include the 
realization of meta-materials which exhibit phenomena like cloaking, transformation optics, negative index refraction, 
etc. The operation frequency for many of these proposals span a wide range from optics \cite{SW11} and micro-
waves \cite{CI06} to acoustics \cite{D13}. Quite recently, a new type of synthetic structure which possesses spatio-
temporal reflection symmetry, or parity-time (${\cal PT}$) symmetry, has emerged. These structures are implemented 
using judicious manipulation of loss and gain mechanisms. The mathematical formalism that describes these systems 
are intrinsically non-Hermitian while the resulting Hamiltonians commute with the joint ${\cal PT}$ operator 
\cite{BB98,BBM99}. Their spectra undergo a transition from real to complex once the parameter 
that controls the degree of non-Hermiticity of the system reaches a critical value. The transition point shows the 
characteristic features of an {\it exceptional point} (EP), where both eigenfrequencies and normal modes coalesce. 
For values of the non-hermiticity parameter which are smaller than the critical value the eigenvectors of the non-
Hermitian Hamiltonian are also eigenvectors of the ${\cal PT}$ operator while above the critical value, they cease to be 
eigenvectors of the ${\cal PT}$ operator. The former domain is termed the {\it exact phase} while the latter the 
{\it broken phase} \cite{BBM99}.

The resulting wave structures show several intriguing features such as power oscillations \cite{MGCM08,RMGCSK10,LSEK12,R12,
ZCFK10}, non-reciprocity of wave propagation \cite{RKGC10,BFB13,Peng14}, unidirectional invisibility \cite{LREKCC11,
R12,FXF13,Lin12} and coherent perfect absorbers and lasers \cite{L10b} etc. Experimental realizations have 
been reported in the framework of optics \cite{R12,Peng14,FXF13,GSDMVASC09} and electronic circuitry 
\cite{LSEK12,BFB13,Lin12} while the applicability of these ideas has been theoretically demonstrated in Bose-Einstein 
Condenstates \cite{GKN08} and in acoustics \cite{acoustics}.

\begin{figure}
\includegraphics[width=1\columnwidth,keepaspectratio,clip]{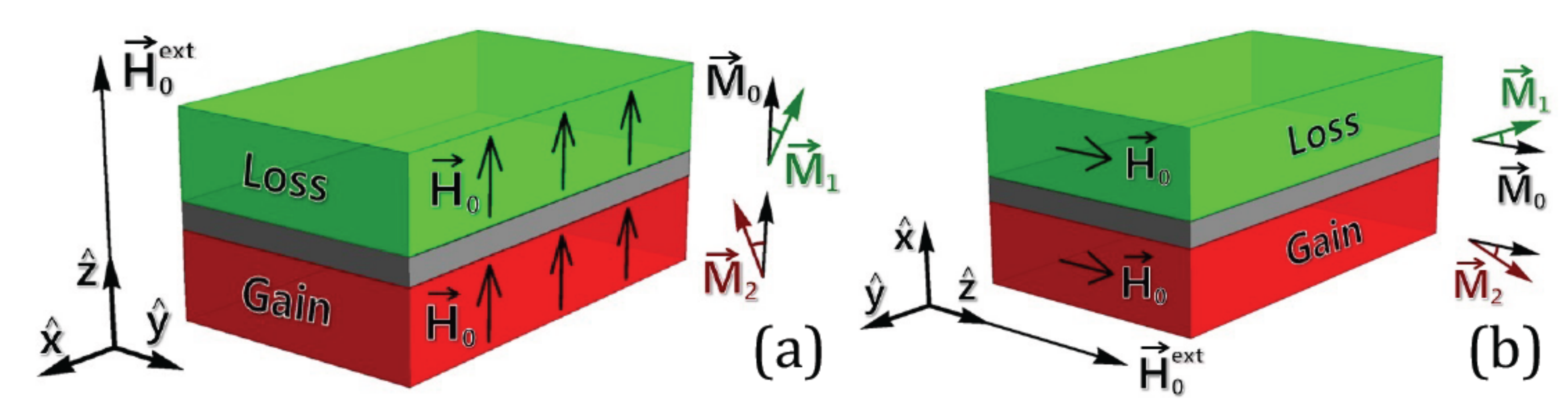}
\caption{Two coupled ferromagnetic films in the presence of an external magnetic field which is along the $z$-axis. We
distinguish between two geometries: (a) Out of plane geometry (the $z$-axis is perpendicular to the films) and (b) In plane 
geometry (the $z$-axis is parallel to the films).
}
\label{fig1}
\end{figure}

Motivated by the success of ${\cal PT}$-symmetric systems, in this paper we propose a class of synthetic magnetic nanostructures 
which utilize natural dissipation (loss) mechanisms together with judiciously balanced amplification (gain) processes in order 
to control magnetization dynamics. Amplification in such structures can be achieved with the help of certain external 
factors such as parametric driving or spin-transfer torque, while loss comes from coupling with phonons or other degrees 
of freedom. As a prototype system we consider two ferromagnetic films (see Fig. \ref{fig1}), one with loss and the other 
with equal amount of gain, coupled by an exchange or by a dipole-dipole interaction. The magnetization dynamics is 
described in terms of two single vector variables, the macroscopic magnetic moments of each film, whose evolution is
given by the non-linear Landau-Lifshitz-Gilbert equations. We will demonstrate that despite the fact that the system is 
non-hermitian, if the gain and loss parameter is below a critical value, the macroscopic magnetic moments precesses 
about the direction of an effective magnetic field inside the sample without being amplified or attenuated. Specifically, 
below a critical value of the gain and loss parameter, the eigenmodes are real while above this critical value, they become 
complex, leading to dynamical instabilities that are limited only by nonlinear effects. The transition point is characterized 
by an exceptional point (EP) degeneracy. Our proposal reveals a new type of non-Hermitian steady state 
dynamics which can be useful for manipulating magnetization switching and potentially lead to new device 
design. Moreover, the realization of EP degeneracies may be utilized for enhanced sensitivity in sensing via frequency splitting.

We consider two ferromagnetic films $n=1,2$ separated by a non-magnetic layer. The two 
geometries that we will consider here are shown in Fig. \ref{fig1}. In Fig. \ref{fig1}a, we assume a uniform external magnetic field 
${\vec H}_{\rm ext}$ perpendicular to the plane of the films (out of plane geometry) while in Fig.~
\ref{fig1}b the external field is parallel to the films (in-plane geometry). The magnetization within each film is uniform
and is represented by a magnetic vector ${\vec M}_{n=1,2}$ with the origin at the center of each film. When the magnetic 
configuration is away from the equilibrium the magnetization precesses around the instantaneous local effective field 
${\vec H}_n$. The latter is generally a complicated function of ${\vec M}_n$ and the external magnetic field ${\vec 
H}_{\rm ext}$. For the cases shown in Fig. \ref{fig1} we have  
\begin{equation}
\label{demagn}
{\vec H}_{n}={\vec H}_{\rm ext} - 4 \pi {\hat N} {\vec M}_n
\end{equation}
where the demagnetizing tensor ${\hat N}$ takes the simple form ${\hat N}_{i,j} = \delta_{i,3}\delta_{j,3}$  for the out of 
plane geometry and ${\hat N}_{i,j} = \delta_{i,1}\delta_{j,1}$ for the in-plane geometry ($i,j=1,2,3$ indicates
the ${\hat x}, {\hat y}, {\hat z}$ directions respectfully). 

The time-evolution of the magnetization dynamics for this coupled system can be described by the following coupled modified 
Landau-Lifshitz (LL) coupled equations:
\begin{eqnarray}
\label{LLG}
{\partial {\vec M}_1\over\partial t}=- \gamma {\vec M}_1 \times {\vec H}_1 -\gamma K {\vec M}_1\times {\vec M}_2+
{\alpha \over \left|{\vec M}_1\right|} {\vec M}_1 \times {\partial {\vec M}_1\over\partial t}\nonumber\\
{\partial {\vec M}_2\over\partial t}= - \gamma {\vec M}_2 \times {\vec H}_2 -\gamma K {\vec M}_2\times {\vec M}_1-
{\alpha \over \left|{\vec M}_2\right|} {\vec M}_2 \times {\partial {\vec M}_2\over\partial t}
\end{eqnarray}
where $\gamma$ is the gyromagnetic ratio. The first term on the right-hand sides of Eqs. (\ref{LLG}) describes the 
interaction of the magnetization ${\vec M}_n$ of each layer with the corresponding local field ${\vec H}_n$. The 
second term represents the coupling between the two ferromagnetic layers. We assume ferromagnetic coupling i.e. $K>0$. 
The last term of the first equation describes dissipation processes and can be introduced in the original LL equations 
by assuming that an effective local friction, proportional to the rate of the change of ${\vec M}_1$, is acting on ${\vec M}_1$, 
pushing it towards the direction of ${\vec H}_1$. It was introduced by Gilbert in order to describe dissipation and can be 
shown to be equivalent to the term that was proposed originally by Landau and Lifshitz for the same purpose \cite{GM96}.
The parameter $\alpha$ is the Gilbert damping term. The last term of the second equation is similar but the sign is reversed, 
reflecting the possibility of amplification mechanisms. We discuss experimentally realizable ways
to achieve "gain" at the suppliment of the paper.

Equations (\ref{LLG}) are invariant under combined parity ${\cal P}$ and time-reversal ${\cal T}$ operations. The
former corresponds to a spatial reflection associated with the change of variables $n=1\leftrightarrow n=2$ while the 
latter corresponds to
a time inversion $t\rightarrow -t$ together with a simultaneous change of the sign of all pseudovectors i.e. ${\vec M}_n
\rightarrow -{\vec M}_n$ and ${\vec H}_n\rightarrow -{\vec H}_n$. This definition of the time-reversal operation is
necessary when magnetic fields, which break the time-reversibility in a Hermitian manner, are 
present. Finally we note that all terms in Eqs. (\ref{LLG}) conserve the length of the magnetization vectors ${\vec M}_n$. 
This can be easily seen by taking the inner product of each of the above equations with the respective 
${\vec M}_n$. This yields ${\vec M}_n {\partial {\vec M}_n\over\partial t} ={1\over 2}{\partial {\vec M}_n^2 \over\partial t}=0$, indicating 
that $\left|{\vec M}_n\right|$ are constants of motion. 

We first analyze the parametric evolution of the eigenfrequencies and normal modes associated with small oscillations 
around the equilibrium configuration as the gain and loss parameter $\alpha$ increases. To this end, we separate the 
magnetization of each film into its equilibrium value, which is assumed to be equal ${\vec M}_n^{(0)}={\vec M}^{(0)}$, 
and its oscillating part ${\vec m}_n$ i.e. ${\vec M}_n={\vec M}^{(0)}+{\vec m}_n$ where $\left|{\vec m}_n\right|\ll 
\left|{\vec M}^{(0)}\right|$. Furthermore, the external
magnetic field can be decomposed into its constant value ${\vec H}_{\rm ext}^{(0)}$ and a time-dependent part ${\vec h}_{
\rm ext}$ i.e. ${\vec H}_{\rm ext}={\vec H}_{\rm ext}^{(0)}+{\vec h}_{\rm ext}$. We mainly focus on the out of plane geometry 
(see Fig. \ref{fig1}a) while at the very end of our presentation we briefly discuss the in-plane geometry (see Fig. \ref{fig1}b).

For the out of plane geometry we recall the relation (\ref{demagn}) which allows us to connect
the external field ${\vec H}_{\rm ext}$ to the local internal field ${\vec H}_n$. Linearizing Eq. (\ref{LLG}) with respect to
${\vec m}_n$ and, furthermore,
setting ${\vec h}_{\rm ext}=0$ we obtain the following linear set of equations
\begin{eqnarray}
{\partial {\vec m}_1\over\partial t}=(\omega_H+\omega_K) {\hat z}\times{\vec m}_1 -\omega_K {\hat z}\times{\vec m}_2
+ \alpha {\hat z} \times {\partial 
{\vec m}_1\over\partial t}\nonumber\\
{\partial {\vec m}_2\over\partial t}=(\omega_H+\omega_K) {\hat z}\times{\vec m}_2 -\omega_K {\hat z}\times{\vec m}_1
- \alpha {\hat z} \times {\partial {\vec m}_2\over\partial t}
\label{linear1}
\end{eqnarray}
where $\omega_K=\gamma K \left|{\vec M}_0\right|$ and $\omega_H=\gamma \left|{\vec H}_0\right|$. Here 
$\left|{\vec H}_0\right|= \left|{\vec H}_{\rm ext}^{(0)}\right|-4\pi \left|{\vec M}_0\right|$ is the constant internal 
magnetic field which is assumed to be the same for both films.

Assuming a harmonic time-dependence for the magnetization ${\vec m}_n(t)={\vec m}_n
\exp(-i\omega t)$, we have
\begin{eqnarray}
-i\omega {\vec m}_1=(\omega_H+\omega_K-i\alpha \omega){\hat z}\times{\vec m}_1 -\omega_K {\hat z}
\times{\vec m}_2\nonumber\\
-i\omega{\vec m}_2=(\omega_H+\omega_K+i\alpha\omega) {\hat z}\times{\vec m}_2 -\omega_K {\hat z}
\times{\vec m}_1
\label{linear2}
\end{eqnarray}
The analysis of Eq. (\ref{linear2}) can be simplified by using the ``center of mass" coordinates of the system. We
define ${\vec \Delta} \equiv {\vec m}_1-{\vec m}_2$ and ${\vec \mu}\equiv {\vec m}_1+{\vec m}_2$.
Then Eqs. (\ref{linear2}) take the following form:
\begin{eqnarray}
\left[(1+\alpha^2)\omega^2-(\omega_H+2\omega_K)^2\right]{\vec \Delta} +2i\alpha \omega(\omega_H+\omega_K){\vec \mu}=0
\nonumber \\
2i\alpha \omega(\omega_H+\omega_K){\vec \Delta} +\left[(1+\alpha^2)\omega^2-\omega_H^2\right]{\vec \mu}=0
\label{linear3}
\end{eqnarray}
which allows us to decouple the $x$ and $y$ components of the center of mass coordinates ${\vec \Delta}, {\vec \mu}$. Thus 
the original set of four coupled equations reduces to two uncoupled sets for the $x$ and $y$ components respectively. 

\begin{figure}
\includegraphics[width=1\columnwidth,keepaspectratio,clip]{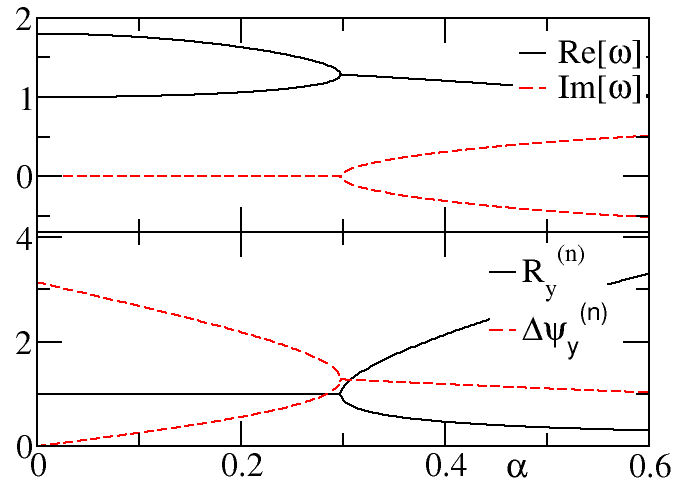}
\caption{(Up) Parametric evolution of the eigen-frequencies of a ${\cal PT}$-symmetric ferromagnetic dimer shown in Fig. \ref{fig1}a.
The parameters used are such that $\omega_K=0.4\omega_H$. (Down) The same but now for the magnitude of the ratio between the
$y$-components of the normal modes and their associated phase difference. The same behaviour holds also for the $x$-components.
}
\label{fig2}
\end{figure}

The eigenvalues and the normal modes can be found by solving the $2\times 2$ secular equation for one of
these components. The eigenfrequencies are given by:
\begin{equation}
\omega_{1,2} = \frac{\omega_H + \omega_K \pm \sqrt{\omega_K^2 - \alpha^2 \omega_H (\omega_H + 2 \omega_K)}}{1+\alpha^2}
\label{freq}
\end{equation}
The limiting case of $\alpha=0$ results in two eigenfrequencies:  (a) $\omega_1=\omega_H$ associated with the ``soft" mode 
(frequency approaches zero when $\left|{\vec H}_0\right|\rightarrow 0$), with ${\vec m}_1={\vec m}_2$ and (b) $\omega_2=
\omega_H + 2 \omega_K$ associated with the "hard" mode, with 
${\vec m}_1=-{\vec m}_2$. As the gain and loss parameter $\alpha$ increases the two eigenfrequencies approach one another
(see Fig. \ref{fig2}) and at some critical value $\alpha=\alpha_{\rm cr}$ they undergo a level crossing and bifurcate into the complex
plane. Using Eq. (\ref{freq}) we calculate the critical frequency $\omega_{\rm cr}$ and the critical value of gain and loss parameter
to be
\begin{equation}
\label{EP}
\alpha_{\rm cr} = {\omega_K\over \sqrt{\omega_H(\omega_H+2\omega_K)}},\quad
\omega_{\rm cr}={\omega_H(\omega_H+2\omega_K)\over \omega_H+\omega_K}
\end{equation}
Near the phase transition point $\alpha_{\rm cr}$, the eigenfrequencies display the characteristic behavior of an exceptional point
$\left|\omega\right|\propto\sqrt{\alpha-\alpha_{\rm cr}}$. This behavior can be exploited in sensing technologies since it 
enhances the sensitivity of frequency splitting detection (for an optics proposal see Ref \cite{W14}).

Next we evaluate the normal modes of the ferromagnetic dimer. Using Eqs. (\ref{linear3},\ref{freq}) we first evaluate ${\vec \Delta}, 
{\vec \mu}$ and from there extract the original variables ${\vec m}_n$. This yields
\begin{equation}
\left(\begin{array}{c}
m_{1x}^{(l)}\\
m_{1y}^{(l)}\\
m_{2x}^{(l)}\\
m_{2y}^{(l)}
\end{array}
\right)
=\left(
\begin{array}{c}
{\alpha (\omega_H+\omega_K) \pm i\sqrt{\omega_K^2 - \alpha^2 \omega_H (\omega_H + 2 \omega_K)}\over(1+i\alpha)\omega_K} \\
\frac{i\big(\alpha (\omega_H+\omega_K) \pm i\sqrt{\omega_K^2 - \alpha^2 \omega_H (\omega_H + 2 \omega_K)}\big)}{(1+i\alpha)\omega_K} \\
-i \\
1
\end{array}
\right)
\label{evec}
\end{equation}
where the sub-indexes$x,y$ refer to the $x,y$ components of the magnetization vectors and the super-index $l=1,2$ refers to the normal mode
corresponding to $+,-$ signs at the rhs of Eq. (\ref{evec}) respectively.
${\cal PT}$-symmetric considerations require that in the exact phase, in contrast to the broken one, these vectors are also eigenvectors 
of the ${\cal PT}$-symmetric operator. In other words, the ratio of the magnitudes of the relevant components $R_{x}^{(l)}\equiv \left| 
{m_{1x}^{(l)}\over m_{2x}^{(l)}}\right| ; R_{y}^{(l)}\equiv \left|{m_{1y}^{(l)}\over m_{2y}^{(l)}}\right|$ in the exact phase is unity indicating
that the magnitude of the magnetization eigenvectors is the same in both the loss and the gain side of the dimer. As $\alpha$ becomes
larger than $\alpha_{\rm cr}$ the magnitude of the magnetization in the loss and in the gain sides become unequal indicating that the 
magnetization eigenmodes reside either on the gain or the lossy side of the dimer. This behavior can be seen nicely in Fig. \ref{fig2}b
where we are plotting $R_{y}^{(l=1,2)}$ as well as the relative phase difference $\Delta \psi_y^{(l=1,2)}$ between the $y$ components 
of the $l=1,2$ modes. We see that for $\alpha=0$ the 
phase difference assumes the values  $\Delta \psi_y^{(l=1)}=0$ and $\Delta \psi_y^{(l=2)}=\pi$ indicating  a symmetric (${\vec m}_1=
{\vec m}_2$ corresponding to the soft mode) and anti-symmetric (${\vec m}_1=-{\vec m}_2$ corresponding to the hard mode) combinations.
At $\alpha=\alpha_{\rm cr}$ we have a degeneracy of the eigenvectors. 

The ${\cal PT}$-symmetric nature of the dimer is also encoded in the time evolution of the magnetization
vectors and the realization of new types of steady-states. The precession dynamics is better represented in 
spherical coordinates i.e. $M_{nx}=M_0 \sin(\Theta_n) \cos(\Phi_n); M_{ny}=M_0 \sin(\Theta_n) \sin(\Phi_n); M_{nz}=
M_0\cos(\Theta_n)$. Specifically we concentrate on the temporal evolution of the polar angle $\Theta_n(t)$, with respect to the direction 
of the internal magnetic fields ${\vec H}_n$. In the absence of gain and loss mechanisms $M_{nz}$
remains constant representing precession around the ${\hat z}$-direction with a fixed angle $\Theta$. When dissipative mechanisms
are taken into account, $\Theta_n$ decreases due to energy losses so that the magnetization vectors align with the 
${\hat z}$-direction. Conversely, in the presence of amplification mechanisms, the magnetization
is driven away from the ${\hat z}$-direction.

\begin{figure}
\includegraphics[width=1\columnwidth,keepaspectratio,clip]{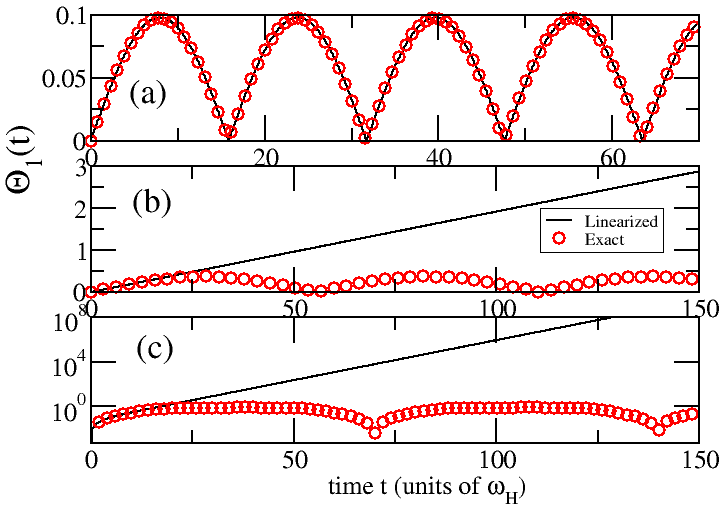}
\caption{Time dependence of the polar angle $\Theta_1(t)$ associated with the magnetization vector of lossy film. The same 
qualitative behavior is exhibited by $\Theta_2(t)$ (not shown here). The initial conditions in all cases are $\Theta_1(t=0)=0$ 
and $\Theta_2(t=0)=0.05$ while $\omega_K=0.4\omega_H$. The results of the exact dynamics Eq. (\ref{LLG}) are indicated with 
red circles while the dynamics generated by the linearized Eqs (\ref{linear1}) are indicated with a black line. (a) Exact phase for
$\alpha=0.85 \alpha_{\rm cr}$; (b) Dynamics at the exceptional point i.e. $\alpha=\alpha_{\rm cr}$; (c) Broken phase with
$\alpha=1.1\alpha_{\rm cr}$. Time is measured in units of inverse $\omega_H$.
}
\label{fig3}
\end{figure}

In the case of ${\cal PT}$-symmetric configurations a different scenario occurs. When $\alpha<\alpha_{\rm cr}$ (exact phase,
see Fig. \ref{fig3}a), despite the fact that the dimer is non-Hermitian, the polar angles $\Theta_n$ oscillate around the initial 
misalignment from the ${\hat z}$-axis without being amplified or attenuated, indicating the existence of a new type of 
steady state. In this domain 
the linearized equations (\ref{linear1}) describe well the exact dynamics (\ref{LLG}). In the broken phase $\alpha>\alpha_{\rm cr}$ (see Fig.
\ref{fig3}c), the evolution generated by the linearized equations (\ref{linear1}) indicates an exponential growth of $\Theta_n$ which 
is associated with the fact that the eigenfrequencies are acquiring an imaginary part. This exponential growth is eventually 
suppressed by non-linear effects which are inherent in the original LL equations (\ref{LLG}). The same behavior is observed
at the phase transition point corresponding to $\alpha=\alpha_{\rm cr}$, with the alteration that the linearized equations (\ref{linear1})
lead to a linear growth of the polar angles $\Theta_n$ (see Fig. \ref{fig3}b). This behaviour is a consequence of the EP degeneracy which results in 
defective eigenmodes.

\begin{figure}
\includegraphics[width=1\columnwidth,keepaspectratio,clip]{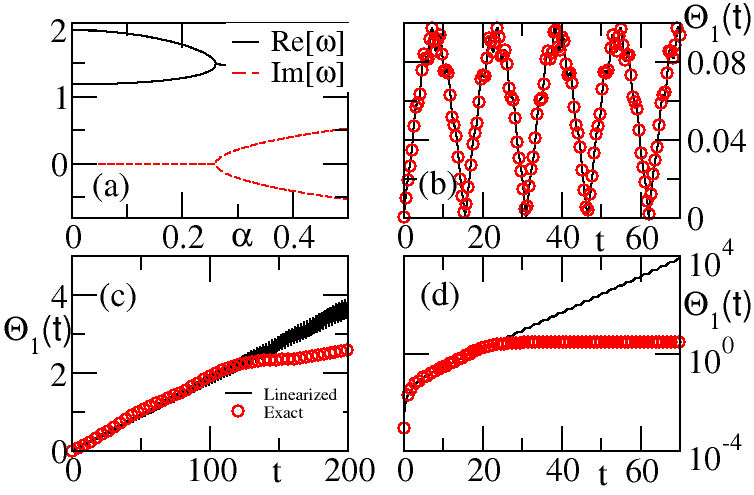}
\caption{In an in-plane geometry (Fig. \ref{fig1}b) for $\omega_K=0.4\omega_H$ and $\omega_M=0.4\omega_H$: 
(a) The parametric evolution of eigen-frequencies versus the gain-loss parameter $\alpha$; The temporal evolution of $\Theta_1(t)$ in the
(b) exact phase with $\alpha=0.85\alpha_{\rm cr}$; (c) EP with $\alpha=\alpha_{\rm cr}$ and (d) broken phase with $\alpha=1.1\alpha_{\rm cr}$.
The initial condition and lines/symbols are the same as in Fig. \ref{fig3}. Time is measured in units of inverse $\omega_H$.
}
\label{fig4}
\end{figure}

For completeness of our study we also analyze the in-plane geometry shown in Fig. \ref{fig1}b. Following the same program as 
previously we can calculate the linearized LL equations (under the condition ${\vec h}_{\rm ext}=0$) that describe the dynamics 
of the magnetization vectors ${\vec m}_n$. For this geometry the equations for ${\vec m}_{n}$ differ from Eq. (\ref{linear1}) by 
an additional term $\omega_M m_n(x) {\hat y}$ at the right hand side where $\omega_M=4\pi\gamma M_0$. The normal modes 
of the system can be calculated assuming an oscillatory behavior of the magnetization vectors while the dynamics can be numerically 
evaluated using the original Eqs. (\ref{LLG}) for the case of in-plane geometry or their linearized version. Some representative results 
for the in plane geometry are reported in Fig. \ref{fig4} showing a qualitatively similar behavior as the out of plane configuration.

In conclusion, we have introduced the notion of ${\cal PT}$-symmetry in magnetic nanostructures. Using 
two coupled ferromagnetic layers, one with loss and another with equal amount of gain, we demonstrated the emergence of a 
new type of steady-state dynamics where the polar angle, although 
not a constant of motion, is bounded and neither attenuates (as in the case of losses) nor amplifies (as in the case of gain). This 
non-Hermitian steady state can be reached for values of the gain and loss parameter $\alpha$ that are below a critical value 
$\alpha_{\rm cr}$. At $\alpha=\alpha_{\rm cr}$ the system experiences an exceptional point degeneracy where both eigenvalues 
and eigenvectors are simultaneously 
degenerate. It will be interesting to extend this study to the case of spin waves (magnons) and investigate the possibility of observing
phenomena such as magnonic Coherent Perfect Absorbers/Lasing, invisibility etc \cite{LREKCC11,L10b}.


\begin{acknowledgments}
We are grateful to V. Vardeni for attracting our interest on the subject. This research was partially supported by an AFOSR MURI grant FA9550-14-1-0037 and by an NSF ECCS-1128571 and 
DMR-1306984 grants. (B.S) acknowledges a Global Initiative grant from Wesleyan University (Dean's office).
\end{acknowledgments}


\end{document}


\title{Supplemental Material}
\maketitle

We point out two possible ways to achieve amplification (gain) of the magnetic oscillations
in ferromagnets.

\subsection{Parametric driving}

Let us first recall the phenomenon of the paramagnetic resonance of a harmonic oscillator \cite{LL}. 
Consider an oscillator whose eigenfrequency is modulated in time so that the equation of motion is
\begin{equation}
\label{po}
{\ddot x}(t)+ \omega_0^2\left[1+\eta \cos(2\omega_0 t)\right] x(t)=0;\quad (\eta\ll 1)
\tag{$S1$}
\end{equation}
The approximate solution of this equation is
\begin{equation}
\label{approx}
x(t)=a(t) \sin(\omega_0 t) + b(t) \cos(\omega_0 t)
\tag{$S2$}
\end{equation}
where the slowly varying amplitudes $a(t), b(t)$ grow exponentially with time, with an increment
($\eta \omega_0/4)=\lambda\ll \omega_0$. Thus, the parametrically driven oscillator exhibits 
an instability (gain). The equilibrium solution $x(t)=0$ of Eq. (\ref{po}) is unstable, i.e. an infinitesimal 
deviation from equilibrium results in an exponential growth. This growth, on top of rapid oscillations 
with frequency $\omega_0$, can be modeled by the equation
\begin{equation}
\label{aem}
{\ddot x}(t) - 2 \lambda {\dot x}(t) + \omega_0^2 x(t)=0.
\tag{$S3$}
\end{equation}
This exponential growth $\exp(\lambda t)$, is eventually limited by non-linear effects.

\begin{figure}[tbp]
\begin{center}
\includegraphics[width=3.3in]{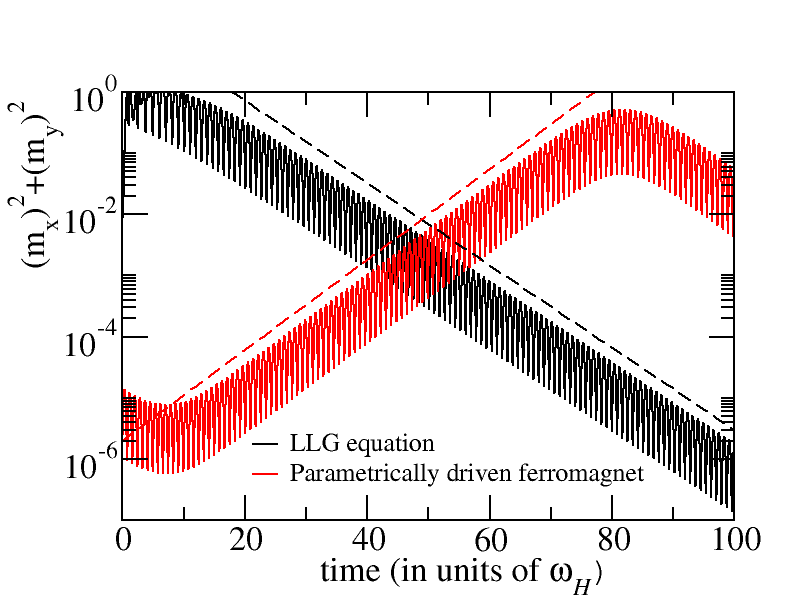}
\end{center}
\caption{The temporal behavior of the magnetization $m(x)^2+m(y)^2$ of a dissipative (black line) 
and an amplified (red) ferromagnet where gain is introduced via parametric driving. The driving parameters 
in the last case are such that the amplification increment has equal magnitude, but opposite sign, with 
respect to the lossy ferromagnet. The dashed lines are drawn in order to guide the eye and indicate 
that the ratios in these two cases are the same. The time is measured in units of inverse $\omega_H$.
}
\label{fig1}
\end{figure}

A similar phenomenon occurs for a magnetic moment driven by an appropriate external magnetic field. 
Consider the in-plane geometry with the external field
\begin{equation}
\label{pofm}
{\vec H}_{\rm ext}=[H_{\rm ext}^{(0)} + h_{\rm ext}(t)] {\hat z}
\tag{$S4$}
\end{equation}
in the ${\hat z}$-direction (in the plane of the film), where the weak, time-dependent component can be written as $H_{\rm ext}^{(0)} 
\eta f(t)$. Since ${\vec h}_{\rm ext}$ is in the same direction as $H_{\rm ext}^{(0)}$ (which is also in the 
direction of the equilibrium magnetization ${\vec M}_0$) it cannot cause the ordinary precession of the 
magnetic moment about the ${\hat z}$-direction. Rather, it can cause an instability via a mechanism 
analogous to the parametric driving of a harmonic oscillator (see Fig. \ref{fig1}). Indeed, neglecting for the moment the losses, 
the linearized Landau-Lifshitz equations read:
\begin{equation}
\begin{array}{lll}
{\dot m}_x &=& -\omega_H \left[1+\eta f(t)\right] m_y\nonumber\\
{\dot m}_y &=& \omega_H \left[1+\eta f(t)\right] m_x+\omega_M m_x\label{pofm1}
\end{array}
\tag{$S5$}
\end{equation}
where $\omega_H=\gamma H_0$, $\omega_M=4\pi\gamma M_0$. (Recall that in this geometry the internal 
field $H_0=H_{\rm ext}^{(0)}$). We do not pursue the detail analysis of Eq. (\ref{pofm1}) but only notice that for 
the case $\omega_M\gg \omega_H$ Eq. (\ref{pofm1})b reduces to ${\dot m}(y)=\omega_M m(x)$ which, after 
taking a time derivative and substituting ${\dot m}(x)$ from Eq. (\ref{pofm1})a, yields ${\ddot m}(y)=-\omega_H
\omega_M\left[1+\eta f(t)\right] m(y)$. For $f(t)=\cos(2\omega_0 t)$, with $\omega_0=\sqrt{\omega_H\omega_M}$, 
this coincides with Eq. (\ref{po}) for the parametrically driven oscillator. Thus a magnetic moment, parametrically 
driven with an ac magnetic field, parallel to the constant field Eq. (\ref{pofm}), exhibits an instability, i.e. an 
exponential growth of the precession angle $\Theta$ about the ${\hat z}$-direction, limited only by nonlinearity. 
Such an instability is modeled by reversing the sign of the attenuation term in the Landau-Lifshitz 
(Gilbert) equation.

Finally, the analysis can be extended to include a decay term into the Landau-Lifshitz equations, in a way similar 
to the inclusion of a weak friction into Eq. (\ref{po}) for the oscillator \cite{LL} (see Fig. \ref{fig1}).

\subsection{Spin transfer torque}

\begin{figure}[tbp]
\begin{center}
\includegraphics[width=3.3in]{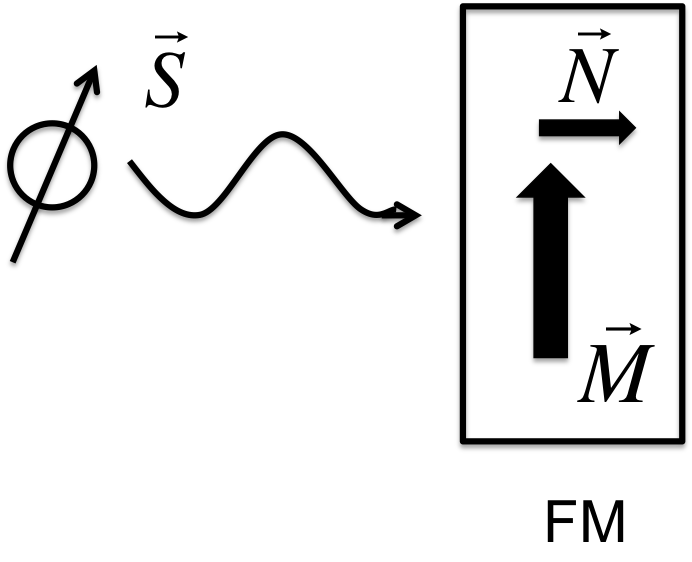}
\end{center}
\caption{A beam of spin-polarized electrons impinges on a ferro-magnetic layer (FM) with magnetic moment ${\vec M}$.
}
\label{fig2}
\end{figure}

A different mechanism for achieving amplification of the magnetic moment precession is based on the spin transfer 
phenomenon (see Ref. \cite{RS07} for a pedagogical review). When spin-polarized electrons are scattered on a ferro-
magnetic layer, they generally transfer some angular momentum to the layer, thus inducing a torque ${\vec N}$ on 
the magnetic moment ${\vec M}$, see Fig. \ref{fig2}. (Spin polarization is usually achieved by passing current through 
another ferromagnetic layer - a "spin polarizer" - not shown in the figure). Two conditions should be satisfied for the 
spin transfer to take place: First, the scattering amplitudes must be spin-dependent, i.e. be different for spin-up 
(parallel to ${\vec M}$) and spin-down electrons (such difference is provided by the exchange splitting between the 
minority and majority spin-bands in the ferromagnet). Second, polarization direction of the incident spins ${\vec S}$, 
should not be strictly parallel to the direction of ${\vec M}$. The angular momentum, transmitted to the ferro-magnetic 
layer by the stream of polarized electrons, affects the dynamics of the magnetic moment ${\vec M}$. The effect is 
described by an amplification term in the Landau-Lifshitz equation. This term has the same form as the damping 
term but with an opposite sign (the resulting equation is referred to as the Landau-Lifshitz-Gilbert-Slonczewski equation).

\begin{figure}[tbp]
\begin{center}
\includegraphics[width=3.3in]{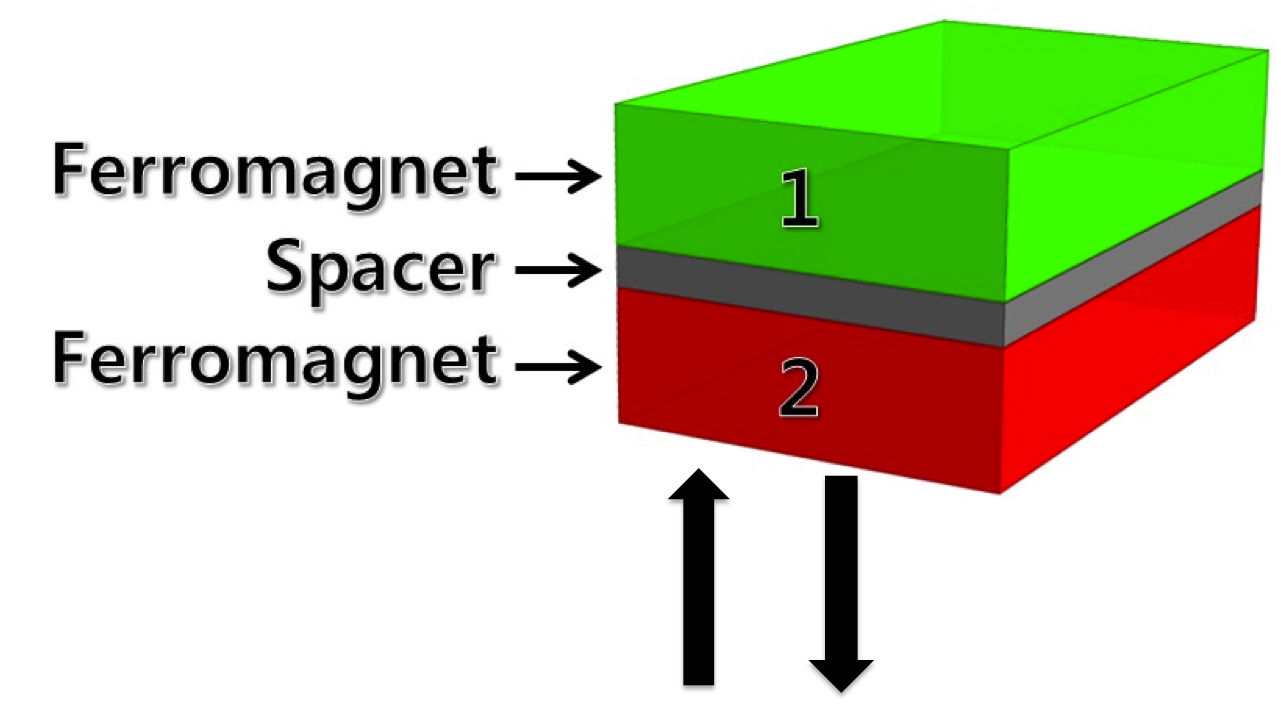}
\end{center}
\caption{Spin-polarized electrons (up-pointing arrow) impinges on the ferromagnetic layer and are reflected back (down-pointing arrow). Spin angular momentum (but no electric current!) is flowing into the layer, creating gain. The red layer indicate the gain ferromagnet while the green layer indicate the lossy ferromagnet.
}
\label{fig3}
\end{figure}

It is interesting to note that spin transfer can occur even in the case of total reflection, provided that the reflection 
amplitudes for up and down-spins, $r_{\rm up}=\exp(i\phi_{\rm up})$ and $r_{\rm down}=\exp(i\phi_{\rm down})$, 
have different phases, see Eq. (14) in Ref. \cite{RS07}. Although the ''transmitted'' wave in this case is purely evanescent, 
so that no charge current can flow into the layer, the angular momentum transmitted to the layer is not zero \cite{RS07,X02}. 
This might provide the most practical way for producing gain in a ${\cal PT}$-symmetric magnetic structure, see 
Fig. \ref{fig3}. Again, as in Fig. \ref{fig2}, we do not show explicitly the set-up which produces the spin-polarized 
current that impinges on the lower film (gain) of our ${\cal PT}$-symmetric device. One can find the full set-up in Ref. \cite{X02}.

